\documentclass[doublecol]{epl2}

\usepackage{amsmath}
\usepackage{xcolor}
\usepackage{amssymb}
\usepackage{graphicx}
\usepackage[colorlinks,bookmarks=false,citecolor=blue,linkcolor=blue,urlcolor=blue]{hyperref}
\DeclareFontFamily{U}{wncy}{}
\DeclareFontShape{U}{wncy}{m}{n}{<->wncyr10}{}
\DeclareSymbolFont{mcy}{U}{wncy}{m}{n}
\DeclareMathSymbol{\Sh}{\mathord}{mcy}{"58}

\def\Eq#1{\hyperref[#1]{Eq.~(\ref*{#1})}}
\def\Equation#1{\hyperref[#1]{Equation~(\ref*{#1})}}
\let\Im\relax\DeclareMathOperator{\Im}{Im}

\title{Absence of a direct metal to Mott insulator transition\\
for disordered interacting fermions on the Cayley tree}
\shorttitle{Absence of a direct metal - Mott insulator transition on the Cayley tree}

\author{Ankita Chakrabarti\inst{1,2,3}  \and Nicolas Laflorencie\inst{3} \and Bertrand Georgeot\inst{3} \and Cyril Martins\inst{4} \and Gabriel Lemari\'e\inst{2,3,5} }
\shortauthor{Ankita Chakrabarti \etal}

\institute{                    
  \inst{1} Department of Physics, National University of Singapore, 2 Science Drive 3, Singapore 117542, Singapore,\\
  \inst{2} MajuLab, International Joint Research Unit IRL 3654, CNRS, Universit\'e C\^ote d'Azur, Sorbonne Universit\'e, National University of Singapore, Nanyang Technological University, Singapore\\
  \inst{3} Laboratoire de Physique Th\'{e}orique, Universit\'{e} de Toulouse, CNRS, UPS, France\\
  \inst{4} Laboratoire de Chimie et Physique Quantiques (UMR 5626), Universit\'e de Toulouse, CNRS, UPS, France\\
 \inst{5} Centre for Quantum Technologies, National University of Singapore, 117543 Singapore, Singapore
}

\abstract{Interacting fermions in the presence of disorder pose one of the most challenging problems in condensed matter physics, primarily due to the absence of accurate numerical tools. Our investigation delves into the intricate interplay between interaction-induced Mott insulation and disorder-driven Anderson localization in the Hubbard model subjected to a random potential. On the Cayley tree, the application of statistical dynamical mean-field theory proves adept at discerning among a metal and the two distinct insulators, Anderson or Mott.  Our comprehensive analysis, accounting for subtle yet potent finite-size effects and fluctuations, yields a noteworthy finding: in the presence of disorder, we consistently observe an intervening Anderson-localized regime between the metallic and Mott insulator states. This observation intriguingly mirrors scenarios witnessed in dirty Bosons, where an insulating Bose glass phase consistently emerges between the superfluid and Mott phases.}

\begin{document}

\maketitle

\section{Introduction}
In condensed matter physics, the metal-insulator transition amid interactions and disorder remains a profound challenge \cite{lee1985disordered, abrahams201050, dobrosavljevic2012introduction, pollak2013electron}. While the mechanism of Anderson localization in disordered systems is well understood \cite{RevModPhys.80.1355, abrahams201050}, extending it to encompass many-body interactions is daunting. Recent studies have shed some light on the many-body localization problem at high energy in one dimension ($d=1$), see \cite{alet2018many, abanin2019colloquium} for recent reviews.  Regarding low-temperature properties, apart from the $d = 1$ situation, where the (un)stability of the Luttinger liquid against disorder has been thoroughly studied~\cite{Giamarchi_1987}, the problem of interacting electrons in the presence of disorder remains a largely open problem for $d > 1$~\cite{Belitz94,Imada98,dobrosavljevic2019metal}. 

In the same time, Mott physics, where Coulomb repulsion induces an insulating state \cite{Mott_1937,Mott_1949}, has been firmly established, but only in periodic ionic potentials. Key insights into the Mott metal-insulator transition {came from key lattice models} \cite{hubbard1963electron,Gutzwiller_1963,Kanamori_1963,Hubbard_1964}, and the advent of Dynamical Mean-Field Theory (DMFT) \cite{PhysRevLett.62.324, PhysRevB.45.6479,RevModPhys.68.13}.
Moreover, when combined with density functional theory, DMFT has successfully opened the door for first-principle calculations in correlated materials~\cite{PhysRevB.57.6884, Anisimov_1997, Biermann_2014, Held_2007}. In essence, DMFT can be seen as a quantum many-body extension of classical mean-field approaches : it operates under the assumption that the dynamics at a specific lattice site can be comprehended as the interplay between the local degrees of freedom at that site and a self-consistent external bath shaped by all other degrees of freedom across different sites. The lattice problem is thus mapped onto a self-consistent quantum impurity problem. Owing to its local assumption, DMFT is exact only in the clean case in infinite dimensions \cite{RevModPhys.68.13}.

In the attempt to extend DMFT to disordered materials~\cite{PhysRevLett.78.3943, dobrosavljevic1998dynamical}, it was discerned that the conventional framework falls short in capturing Anderson localization~\cite{dobrosavljevic1998dynamical}. This discrepancy arises because the spatial uniformity presumed by DMFT contradicts the essence of Anderson localization, where electronic states are exponentially localized, corresponding to large spatial inhomogeneities \cite{RevModPhys.80.1355}.

Various DMFT extensions have been explored to incorporate localization effects (see~\cite{miranda2012dynamical, byczuk2010anderson} for reviews): {Typical Medium Theory~\cite{dobrosavljevic2010typical} considers the typical (most probable) local density of states as an order parameter \cite{PhysRevB.92.144202}. Cluster DMFT allows resolution of spatial inhomogeneities up to the cluster size~\cite{PhysRevB.92.014209, PhysRevB.92.205111, PhysRevB.95.144208, PhysRevLett.118.106404}.}
Lastly, statistical DMFT accounts for all spatial fluctuations of the bath, ensuring exactness in the non-interacting limit while converging to standard DMFT in the pure, non-disordered scenario~\cite{PhysRevLett.78.3943, dobrosavljevic1998dynamical, PhysRevB.84.115113}.

Here we consider a fundamental problem where statistical DMFT is expected to be highly reliable: the metal-insulator transition driven by interactions and disorder on the Cayley tree (see \cite{PhysRevLett.78.3943, dobrosavljevic1998dynamical, PhysRevB.84.115113, PhysRevLett.110.066401, HOANG2019320} for previous studies). This regular network structure, in which each node has the same number of neighbors, has no loops and mimics infinite effective dimensionality. 
On the Cayley tree, the presence of disorder introduces interesting and non-trivial non-ergodic properties, as recently highlighted in several studies about the Anderson transition in the non-interacting limit \cite{abou1973selfconsistent, zirnbauer1986localization, monthus2008anderson, biroli2012difference, deluca2014anderson, PhysRevLett.118.166801, tikhonov2016fractality, kravtsov2018non, biroli2018delocalization, garcia2019two}. This leads to large fluctuations of local quantities with distributions having fat tails. Addressing these properties is crucial for an accurate description of the metal-insulator transition,  and this can be done with statistical DMFT \cite{PhysRevLett.78.3943, PhysRevB.84.115113, PhysRevLett.110.066401}.

A critical challenge of statistical DMFT lies in the selection of the method to solve the impurity problems. Since a very large number of impurity problems must be solved, a very fast solver is required. In this study, we have chosen the analytical solver based on the ``atomic limit" referred to as ``Hubbard-I approximation" \cite{hubbard1963electron}. 
Due to its simplicity, it remains a first-choice solver to study the Anderson-Hubbard model \cite{PhysRevB.77.054202}.
Moreover, 
this approximation has been shown to hold in the strong disorder limit \cite{Song_2009},
the regime we focus on in this study.

In the non-interacting limit, the Anderson transition on graphs of effectively infinite dimensionality faces intricate finite-size effects \cite{tikhonov2016anderson, pietracaprina2016forward, tarquini2017critical, PhysRevLett.118.166801, tikhonov2019critical, garcia2019two, biroli2018delocalization, Crt_4, sierant2023universality}, making it challenging to pinpoint the critical disorder value, denoted as $W_c^0$. Recent studies have revealed a critical disorder in the range $18.09\le W_c^0 \le 18.18$ \cite{tikhonov2019critical, parisi2019anderson, sierant2023universality}, a value significantly higher than the previous estimate ($W_c^0 \approx 16$) \cite{abou1973selfconsistent}. These findings underscore the profound impact of carefully considering the non-trivial properties of infinite-dimensional graphs, leading to an enlargement of the metallic phase.

With interactions, the determination of the localization transition point becomes even more complex. Previous studies \cite{Alvermann_2006,PhysRevB.76.035111, PhysRevB.76.045105} found that the threshold depends significantly on the observable considered. Consequently, very different metal-insulator phase diagrams have emerged on the Cayley tree \cite{PhysRevLett.78.3943, dobrosavljevic1998dynamical, PhysRevLett.110.066401, PhysRevB.84.115113, HOANG2019320}.

In this study, we introduce two methods that precisely identify the transition point, accounting for the logarithmically slow finite-size effects and the substantial fluctuations arising in this problem. These methods accurately determine $W_c^0$ in the non-interacting limit, aligning with recent precise determinations \cite{tikhonov2019critical, parisi2019anderson, sierant2023universality}. Crucially, our approach extends to the interacting case, enabling to map the disorder- and interaction-driven metal-insulator phase diagram. Our key finding reveals the absence of a direct metal-to-Mott insulator transition on the disordered Cayley tree. We always observe an Anderson insulating phase between the metal and the Mott insulator, a situation reminiscent of the bosonic case~\cite{PhysRevLett.103.140402, PhysRevB.80.214519}.


\section{Model and statistical DMFT method}\label{Sec_1}
In this letter, we focus on investigating the simplest scenario where DMFT should provide an accurate description of the disorder and interaction driven metal-insulator transition \cite{PhysRevLett.78.3943}. To achieve this, we explore the Anderson-Hubbard model, which accounts for the interplay between local Hubbard repulsion and onsite disorder. We  consider this model on the Cayley tree with  Hamiltonian:
\begin{equation}
H = \sum_{\langle ij \rangle} \sum_\sigma \left(-t + (\mu-\varepsilon_i) \delta_{ij}\right) c_{i,\sigma}^\dagger c_{j,\sigma} + U \sum_i n_{i,\uparrow} n_{i,\downarrow},
\label{Hamiltonian}
\end{equation}
where $c_{i,\sigma}^\dagger (c_{i,\sigma})$ describes the creation (annihilation) operator at site $i$ with spin $\sigma=\pm{1}/{2}$. The number operator at each site is $n_{i,\sigma}=c_{i,\sigma}^\dagger c_{i,\sigma}^{\vphantom{\dagger}}$. 
We consider a finite Cayley tree of depth $n$, with a branching number $K$.
The hopping amplitude between a site $i$ and its $(K+1)$ nearest neighbors (indexed by $j$) is given by $t$.
The local Coulomb interaction is given by $U$ and $\mu$ corresponds to the chemical potential. 
The on-site potential $\epsilon_i$ at each site $i$ are 
independent random variables uniform in $[-\frac{W}{2},\frac{W}{2}]$.

We investigate {the metal-insulator phase diagram of the above problem in the regime of large disorder $W$, arbitrary interaction $U$ at half-filling.}
Employing statistical DMFT \cite{PhysRevLett.78.3943,dobrosavljevic1998dynamical,SDMFT_3, PhysRevB.84.115113, PhysRevLett.110.066401}, our study focuses on three key quantities: 1) The impurity Green's function $G_{i}$ at a site $i$ of the network. Its imaginary part provides the local density of states (LDOS), distinguishing between a metal, an Anderson insulator, or a Mott insulator. 2) The self-energy $\Sigma_i$, which quantifies many-body effects in the impurity. 3) The hybridization $\Gamma_i$, characterizing the bath coupled to the impurity. In the "standard" DMFT, the Green's function, self-energy and hybridization are interrelated through a self-consistent relation; within statistical DMFT  considered here, they are interrelated via a recursive relation.

The statistical DMFT approach is designed to yield exact results in the absence of interactions. In the context of the Anderson localization problem on the Cayley tree, an exact solution can be obtained using the cavity method \cite{abou1973selfconsistent, PhysRevLett.78.3943, FIM:PRB10, biroli2010anderson, biroli2012difference}. This involves creating a cavity at the neighboring site $j$ of the impurity $i$, with $j$ being closer to the root of the Cayley tree. Considering the cavity Green's function of the impurity $G^{(j)}_i$, it can be demonstrated that $G^{(j)}_i = (E - \varepsilon_i - t^2 \sum_{k=1}^K G^{(i)}_k)^{-1}$, with $E$ the energy \cite{abou1973selfconsistent}. Here, the summation is over the $K$ neighbors of $i$, indexed by $k$, that are farther from the root. Importantly, for these $k$ sites, the cavity is at site $i$, making them disconnected: the $K$ branches become entirely independent and can be seen as an effective bath for the impurity problem at $i$ \cite{PhysRevLett.78.3943}. Consequently, 
$\Gamma_i\equiv t^2 \sum_{k=1}^K G^{(i)}_k$
represents the hybridization at $i$. Due to the presence of disorder, translational invariance is broken, leading to distinct impurity problems at each site $i$ coupled to a different effective bath. One can solve this problem recursively, going from the leaves to the root \cite{monthus2008anderson, FIM:PRB10, lemarie2013universal, biroli2018delocalization, chakrabarti2022traveling}.

In the presence of interactions, incorporating a self-energy $\Sigma_i$ into the cavity Green's function allows us to describe Hubbard interaction effects  via
$
    G^{(j)}_i = (E - \varepsilon_i - \Sigma_i - \Gamma_i)^{-1}
$ \cite{RevModPhys.68.13}.
This leads to an Anderson impurity problem \cite{anderson1961localized} 
solved using a quantum impurity solver. In our case, solving quantum impurity problems corresponds to the number of sites in the finite Cayley tree. Due to anticipated finite-size effects, a deep Cayley tree is necessary (number of sites $N$ up to $\approx 4 \cdot 10^6$), demanding a fast solver. We use the Hubbard-I approximation \cite{hubbard1963electron},
which gives the following analytical expression:
\begin{equation}
\Sigma_{i}(E) = \frac{Un_{i}}{2}  + 
\frac{U^2\frac{n_{i}}{2} \left(1- \frac{n_{i}}{2}\right)}{E+\mu-\varepsilon_{i}-U(1- \frac{n_{i}}{2})} \;,
\label{HI}
\end{equation}
where $n_i$ is the occupation number at site $i$, at zero temperature and assuming a paramagnetic state.

We solve this problem iteratively from the boundary to the root of the Cayley tree following a similar method as in Ref.~\cite{chakrabarti2022traveling}. Our main focus is the local density of states (LDOS) at the root, $\rho \equiv \frac{1}{\pi}\Im G^{(r)}_{j}$,
where $j$ represents one of the $K+1$ nearest neighbors of the root $r$. 
Specifically, this quantity captures the system's response at the root to a small imaginary shift in energy  $\eta$: $E \rightarrow E-i \eta$. We examine $\rho$ at the Fermi energy $E=0$, and study the behavior of its distribution, $P(\rho)$, with increasing tree depth $n$ 
. The LDOS distinguishes our system's phases \cite{abou1973selfconsistent, PhysRevLett.78.3943}: the Mott insulator displays a gap at $E=0$ ($\overline{\rho}(E=0) \rightarrow 0$, where $\overline{(\ldots)}$ represents disorder averaging), while both the metal and the Anderson insulator are gapless compressible ($\overline{\rho}(E=0) > 0$). However one can distinguish the Anderson insulator, where the typical value $\rho^{typ}=\exp\left({\overline{\ln \rho}}\right)$ vanishes, indicating substantial fluctuations ($\overline{\rho} \gg \rho^{typ}$), from the metallic phase which exhibits finite $\rho^{typ}$.

We perform numerical simulations on a Cayley tree with $K=2$ and depth $n=6-22$.
For each value of $n$, $W$ and $U$,  $10^{4}$ disorder realizations are considered.
We choose $\mu = \frac{U}{2}$ corresponding to half-filling {$\frac{1}{N}\sum_{i}n_{i} = 1$}. 
We set the hopping amplitude to $t=1$ and focus on the regime of large disorder $W$ of the order of $W_c^{0}$.

\section{Phase diagram}\label{Sec_2}
\begin{figure}
\centering
		\includegraphics[angle=0,width=\linewidth]{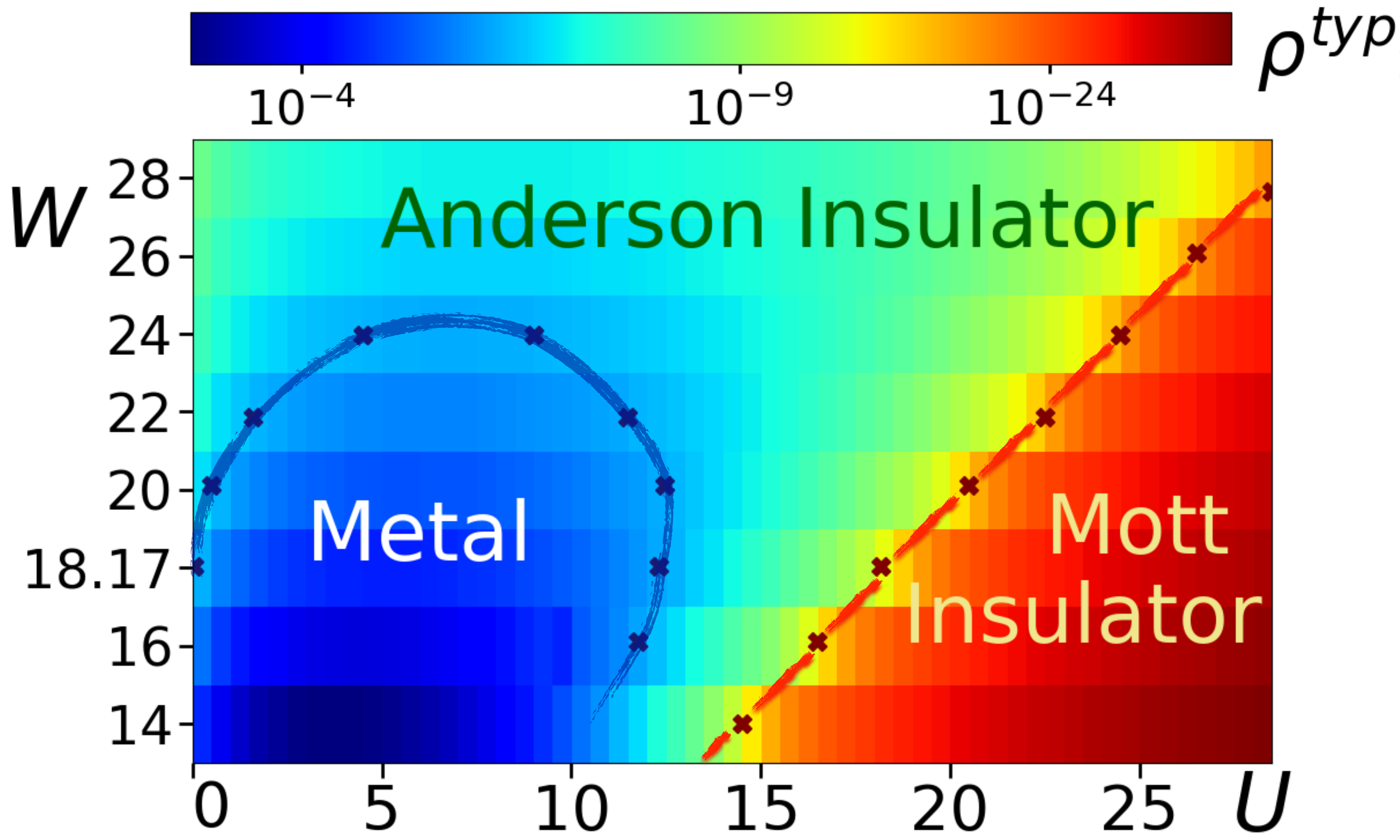}
  \caption{The Metal-Insulator phase diagram of the Anderson-Hubbard model on the Cayley tree at half-filling and in the large disorder regime ($W\in [14,28]$). 
  Colors represent the typical value of the LDOS at the root ($\rho_{typ}$) with a linear color scale for $\ln (- \ln \rho_{typ})$. Three distinct phases are visible: Metal (finite $\rho_{typ}$, in blue), Anderson Insulator (cyan), and Mott Insulator (red). 
  The data corresponds to $\eta=10^{-2}$ at the boundary and a total of $n=16$ generations. Critical interaction strengths ($U$) at the localization transition (blue crosses; blue line is a guide to the eye) are precisely determined using the traveling wave criteria discussed in the main text and Fig.~\ref{Fig_3}. On the right side ($W\approx U$), the transition (red crosses, red line  is a guide to the eye) between the Anderson Insulator and the Mott Insulator corresponds to the opening of a Mott gap at the Fermi energy. Notably, a direct metal-Mott transition is never observed.}
  \label{Fig_PD}
 \end{figure}

We first present the metal-insulator phase diagram obtained from our study of the Anderson-Hubbard model Eq.~\eqref{Hamiltonian} on the Cayley tree in Fig.~\ref{Fig_PD}. 
Three distinct phases emerge: Metal, Anderson insulator, and Mott insulator.
Critical interaction values for the localization transition are determined through the new criteria we introduce in the following section, see Fig.~\ref{Fig_3}. 
The line separating the Anderson and Mott insulating phases, shown on the right side of the phase diagram, corresponds to the opening of a Mott gap at the Fermi energy, investigated later.

A significant deviation from previous studies \cite{PhysRevB.84.115113, PhysRevLett.110.066401} is evident.
As $U$ increases at fixed disorder, an intermediate Anderson-localized phase appears before entering the Mott insulator. Notably, there is no direct transition from the metal to the Mott phase in the regime of large disorder we have considered. It is important to emphasize that this observation contradicts the conventional expectation: the subtle finite-size effects and substantial fluctuations that we elaborate on in the rest of the Letter typically enlarge the metallic phase, pushing the critical disorder in the non-interacting limit from $W_c^0\approx 16$ to $W_c^0\approx 18.17$. Consequently, these factors should have promoted the metallic regime, thereby not challenging the direct metal-Mott transition observed in previous studies  \cite{PhysRevB.84.115113, PhysRevLett.110.066401}.


\section{Accurate determination of the localization threshold}\label{Sec_3}

Here we present an accurate method for locating the threshold between the metal and the Anderson localized phase. In the absence of interactions, the Anderson transition on networks of effective infinite dimensionality, such as the Cayley tree, faces significant and non-trivial finite-size effects \cite{tikhonov2016anderson, pietracaprina2016forward, tarquini2017critical, PhysRevLett.118.166801, tikhonov2019critical, garcia2019two, biroli2018delocalization, Crt_4, sierant2023universality}. We establish an analogy between localization and traveling wave transitions in a branching random walk with an absorbing wall \cite{TW_2, TW_1, monthus2008anderson, chakrabarti2022traveling}, allowing us to leverage known analytical results from this problem. This approach enables us to precisely determine (i) the critical disorder for the Anderson transition at zero interaction strength, confirming recent predictions \cite{tikhonov2019critical, parisi2019anderson, sierant2023universality}, and (ii) extend the method to get the critical interaction strengths $U_c$ for various fixed disorder values $W$, thus mapping the phase diagram Fig.~\ref{Fig_PD}.

\begin{figure}
\centering
	 \includegraphics[angle=0,width=0.45\linewidth]{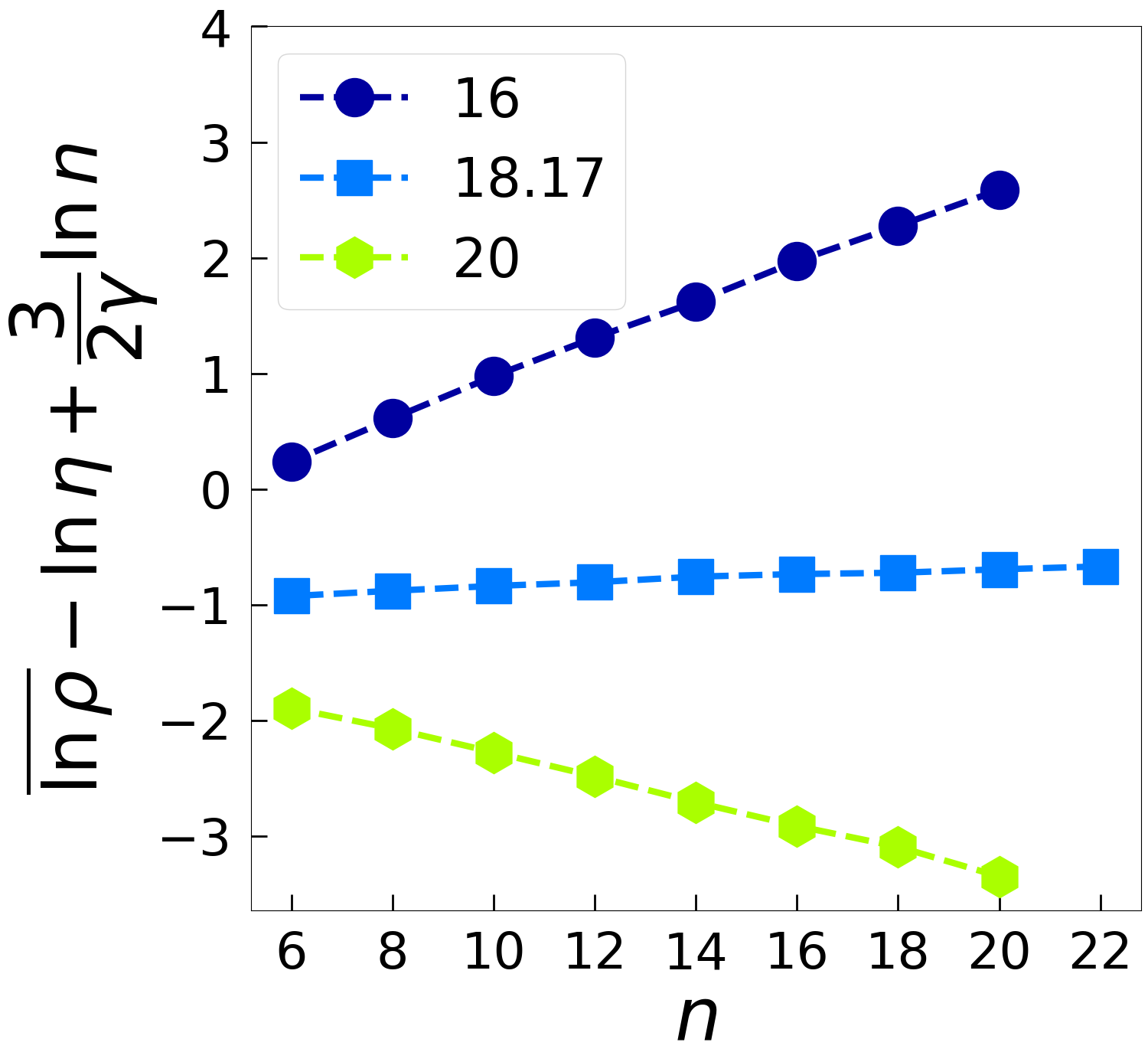}
      \includegraphics[angle=0,width=0.45\linewidth]{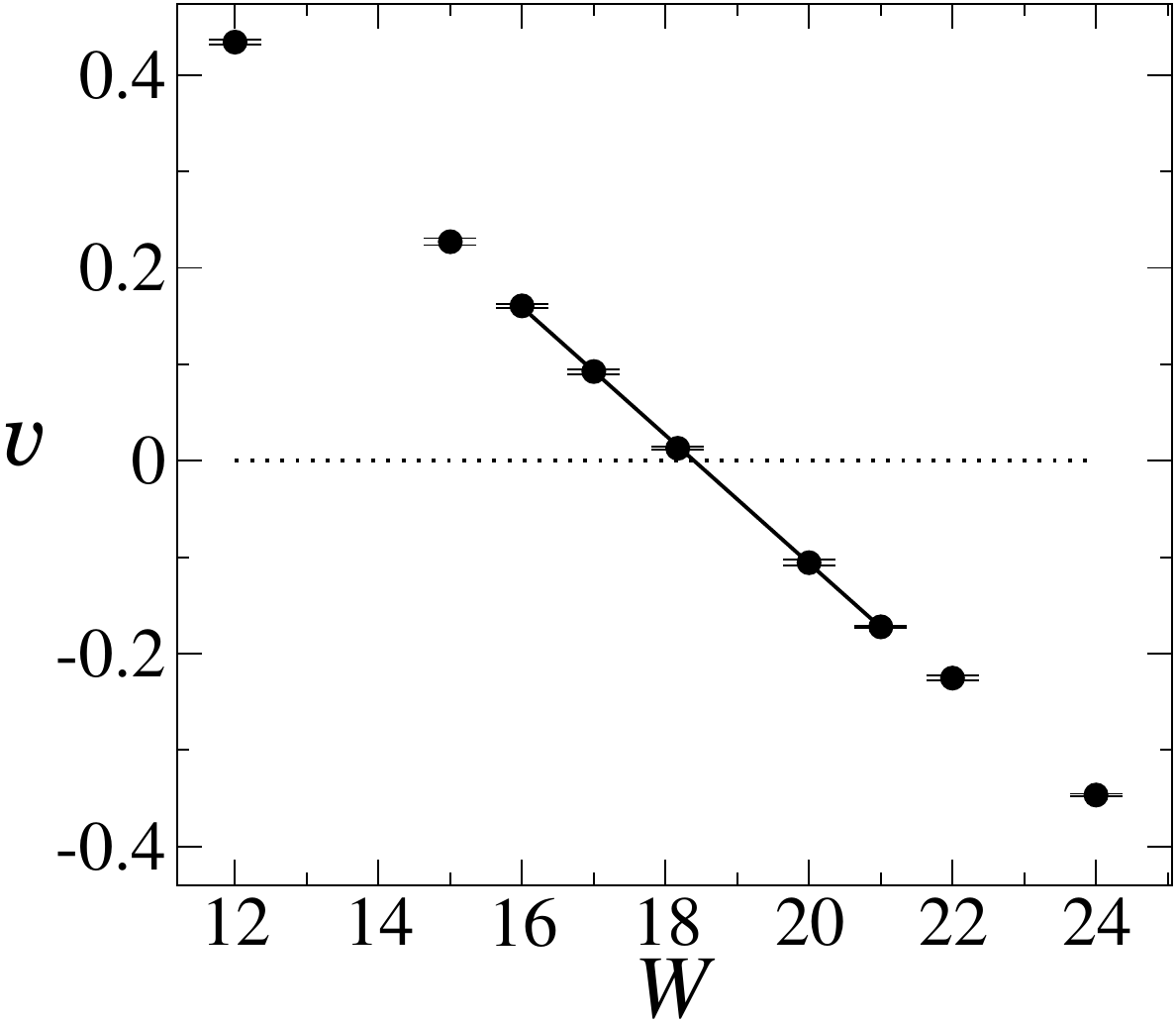}\\	
      \includegraphics[angle=0,width=\linewidth]{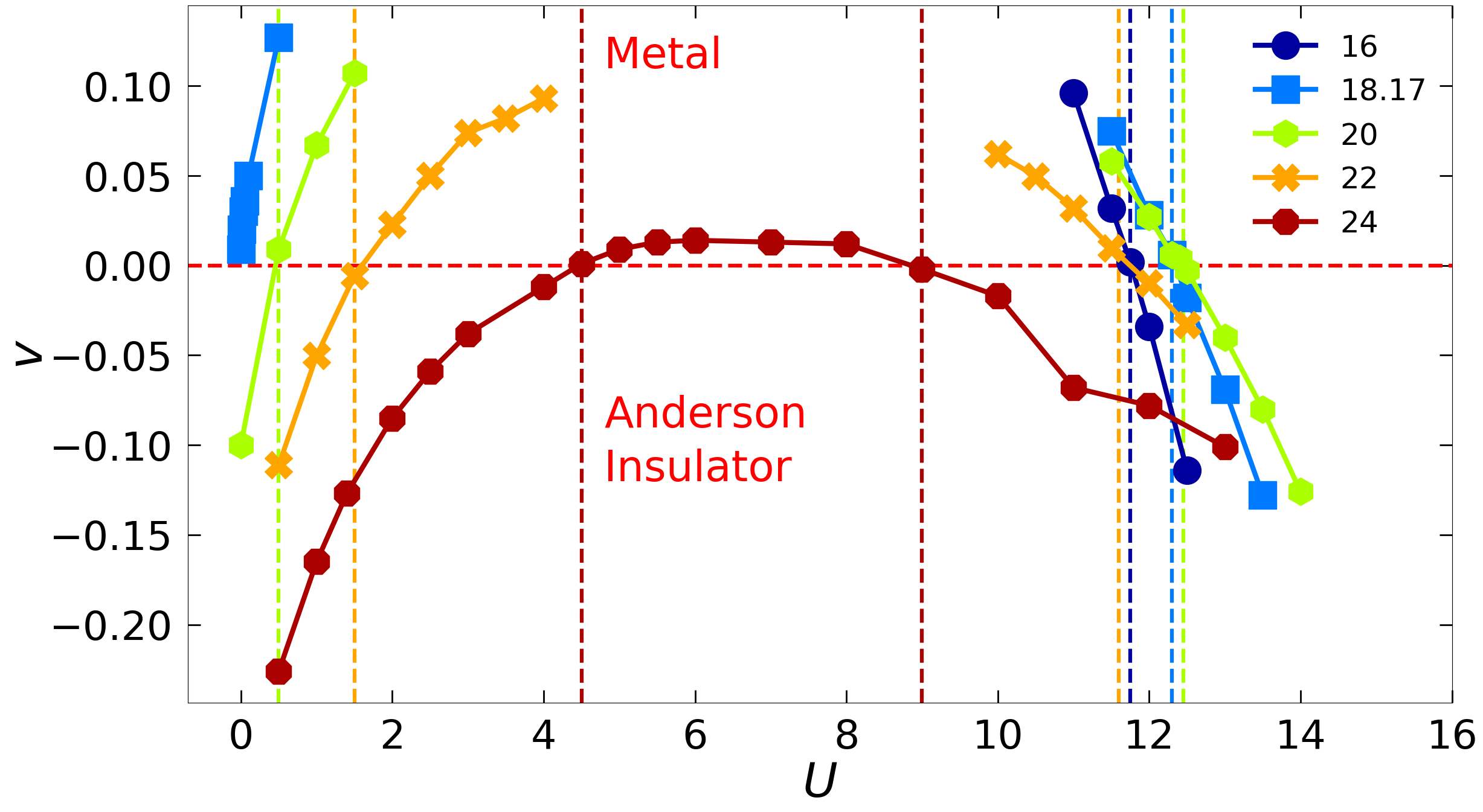}
  \caption{Accurate determination of the localization transition threshold by analyzing the traveling wave behavior of $\rho$. Applying a small $\eta=10^{-8}$ at the boundary leads to a traveling wave pattern in $\mathcal P(\ln \rho)$ with a velocity $v$, determined using Eq.~(\ref{TW_eqn}) with $\gamma=1/2$, that becomes zero at the transition point. Upper panels: non-interacting case. Left: In the delocalized case $W=16$, $\overline {\ln \rho} - \frac{3}{2 \gamma} \ln n $ increases linearly with $n$ with $v>0$, while $v<0$ in the localized phase, and $v=0$ at the transition point. We find $W_c^0= 18.39\pm0.22$ as indicated by the linear fit of $v$ vs. $W$ shown in the upper right panel, a value in good agreement with previous studies \cite{tikhonov2019critical, parisi2019anderson, sierant2023universality}. 
  Lower panel: Interacting case at various disorder values
  $W=16,18.17,20,22,24$ (indicated by different colors). At $W=16$, a transition from a metallic ($v>0$) to a localized phase ($v<0$) occurs at $U_c=11.6$. Conversely, for $W=20,22,24$, delocalization transitions shift from negative to positive velocity at the onset of $U$, followed by reverse localization transitions at higher $U$ values. Notably, these values are significantly smaller than the threshold $U \approx W$ of the Mott transition. Thus, a localized (Anderson insulator) regime consistently precedes the Mott phase.}		
  \label{Fig_3}
 \end{figure}

We introduce $\eta$ at the boundary and analyze the response $\rho$ at the root to identify the metal or Anderson insulating phases (the Mott insulating phase will be discussed later). When a significant $\eta=10^{-2}$ is applied at the boundary (as considered in Fig.~\ref{Fig_PD}), we observe a transition between: (a) a traveling wave regime, where the distribution $\mathcal P(\ln\rho)$ propagates with $n$, preserving its shape, to smaller values of $\overline{\ln \rho}$, and (b) a stationary phase, where $\mathcal P(\ln\rho)$ converges to a fixed distribution at large $n$. In the first traveling case, $\rho_{typ}$ vanishes exponentially with $n$: $\ln\rho^{typ} \simeq vn$, where the velocity of the traveling wave $v<0$. This is a characteristic localized behavior, with $v$ being the inverse of the localization length $v= -1/\xi$. On the other hand, the stationary regime corresponds to the delocalized phase, where $\rho_{typ} > 0$.

To accurately determine the transition point, it is crucial to precisely account for finite-size effects that modify these asymptotic behaviors (valid only at large $n$), especially considering the limited range of $n$ from $6$ to $22$ accessible.
The analogy with the traveling wave problem \cite{TW_2, TW_1, monthus2008anderson, chakrabarti2022traveling} proves invaluable here, predicting the slow logarithmic finite-size effects in the traveling phase \cite{TW_1, chakrabarti2022traveling}:
\begin{equation}\label{TW_eqn}
\overline{\ln \rho} = vn -\frac{3}{2 \gamma} \ln n + \text{constant}.
\end{equation}
The significance of this prediction lies in its universality. 
The prefactor of $\ln n$ depends solely on $\gamma$, the characteristic tail exponent of $P(\rho) \sim \rho^{-(1+\gamma)}$ at large $\rho \gg \rho_{typ}$. 
This exponent is a crucial feature of the localization transition, and its critical value is established as $\gamma_c=1/2$ \cite{abou1973selfconsistent, tikhonov2019critical}. This universality enables us to pinpoint the transition precisely, as we will illustrate.

If we introduce an infinitesimal $\eta=10^{-8}$ at the boundary, the behavior typical of the localized phase described in Eq.~(\ref{TW_eqn}) persists even in the delocalized phase, at least for not too large depths $n$ \cite{FIM:PRB10, kravtsov2018non, chakrabarti2022traveling}. In this scenario, $\rho_{typ}$ initially \textit{increases} exponentially before reaching its stationary value at large $n$. The velocity $v$ is thus positive ($v>0$) in the delocalized phase and the localization transition corresponds to $v=0$.
Our initial strategy for pinpointing the transition point involves determining $v$ under the influence of $\eta=10^{-8}$ and incorporating logarithmic corrections with $\gamma=1/2$ as described in Eq.~\eqref{TW_eqn}, see also \cite{TW_1, chakrabarti2022traveling}. We ascertain the non-interacting critical disorder value $W_c^0$ and subsequently establish the value of $U$ at a specific $W$ where $v=0$ \cite{monthus2008anderson, chakrabarti2022traveling}.

We present the numerical results for the non-interacting case in Fig.~\ref{Fig_3}. The typical LDOS $\overline{\ln \rho}$, accounting for appropriate logarithmic corrections as per Eq.~(\ref{TW_eqn}) with $\gamma=1/2$, is plotted against $n$ in Fig.~\ref{Fig_3}(a) for different disorder values $W$ across the transition. The slopes obtained from linear fits yield the corresponding values of the velocity $v$, which are reported in Fig.~\ref{Fig_3}(b) as a function of disorder $W$. We find that $v=0$ corresponds to $W_c^0=18.39 \pm 0.22$,
a result in good agreement with recent studies \cite{tikhonov2019critical, parisi2019anderson, sierant2023universality}.

In the presence of interactions and large disorder ($W\in [16,24]$), we observe distinct velocity behaviors, see Fig.~\ref{Fig_3}(c). For $W=16$, the system starts as a metal ($v>0$) for $U=0$, before getting to a localized phase ($v<0$) at $U_c\approx 11.6$. When $W=18.17 \approx W_c^0$, a delocalized phase ($v>0$) is immediately observed at the onset of $U$, and then there is a transition into a localized phase for $U>U_c\approx 12.3$. When $W>W_c^0$, a localization-delocalization transition is found at small $U$, observed in the change from negative to positive velocity ($U_c\approx 0.5,1.5,4.5$ for $W=20,22,24$). Conversely, a reverse transition from positive to negative velocity occurs again at larger $U$ values, indicating a localized, Anderson insulator, phase before the Mott transition takes place at $U\approx W$ (further discussed later).

 \begin{figure}
\centering
           \includegraphics[angle=0,height=5cm,width=\linewidth]{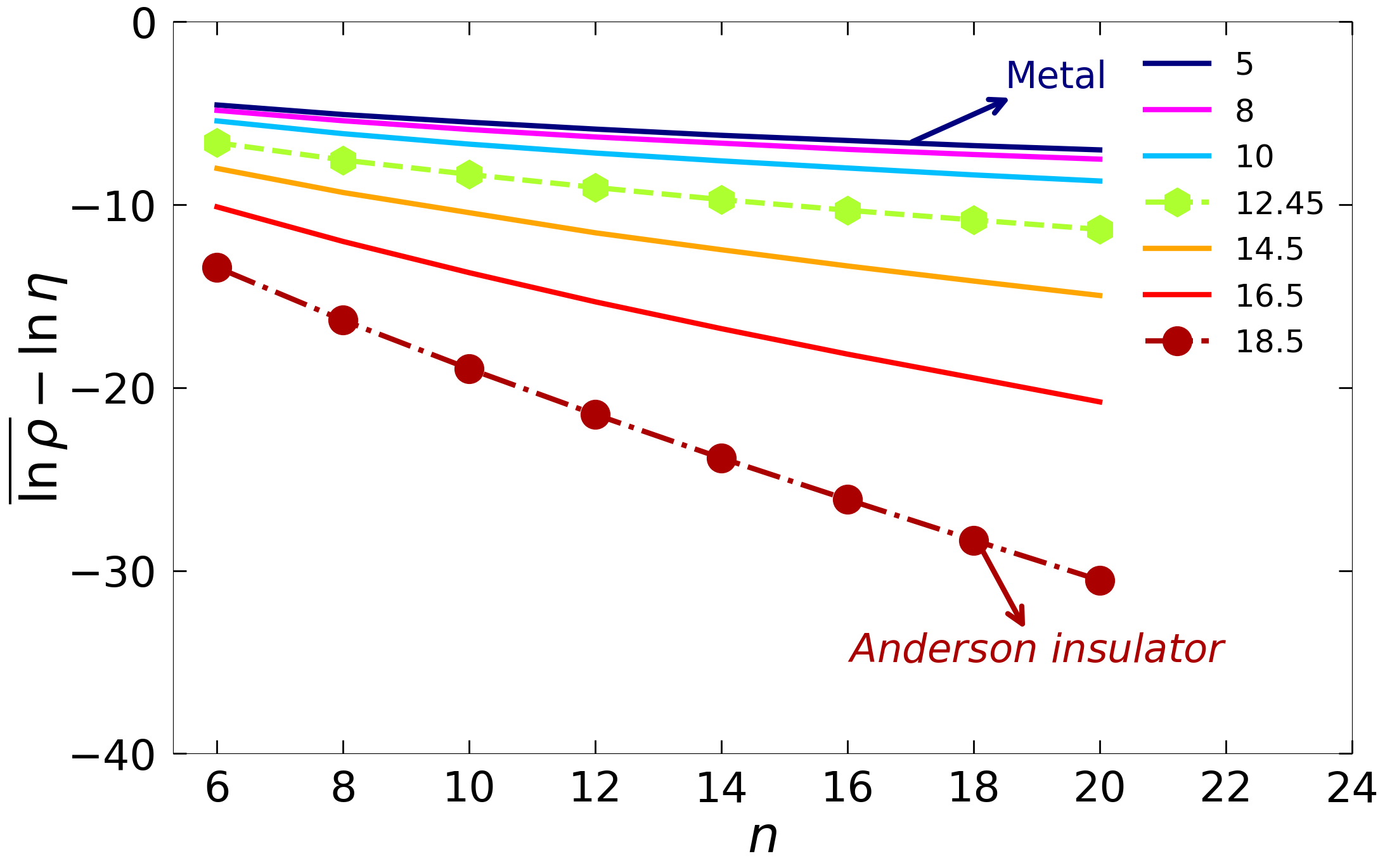} \caption{Critical behavior of $\overline{\ln \rho}$ vs. $n$ at the localization transition for large $W=20$ and $\eta=10^{-2}$ at the boundary. A transition from a stationary regime for $\rho$, clearly observed at $U=5$, to a traveling behavior characterized by Eq.~(\ref{TW_eqn}) with $v \approx -1.21$ at $U=18.5$ (dashed-dotted line) occurs at $U_c=12.45$. The critical behavior is well-fitted by \Eq{eq:critsmallsizecorr} with $C' \approx 4.3$ and $c_1\approx 2.9$ (dashed line). This result corresponds well to our previous prediction of the localization transition threshold shown in Fig.~\ref{Fig_3}.
}.
 \label{Fig_4}
 \end{figure}

 The predicted localization transition thresholds can be verified for the more physically relevant choice $\eta=10^{-2}$, where the transition to a stationary regime for $\mathcal P (\ln \rho)$ becomes effective. Utilizing the analogy with the traveling phase transition \cite{TW_2, chakrabarti2022traveling}, we expect that at the critical point, $\rho^{typ}$ vanishes with $n$ following a stretched exponential form:
\begin{equation}\label{eq:critsmallsizecorr}
\overline{\ln \rho} = - C' ({n} - c_1)^{1/3}\; ,
\end{equation}
where $C'$ and $c_1$ are small size correction constants. We systematically verified that the previously determined transition points at $\eta=10^{-8}$ yield the correct critical behavior at $\eta=10^{-2}$. 

In Fig.~\ref{Fig_4}, we present the behavior of $\overline{ \ln \rho}$ as a function of $n$ for large $W=20$ and interaction strengths ranging from $5$ to $20$, all exceeding the initial localization-delocalization transition at $U_c=0.5$. At $U=5$, a stationary regime is evident, indicating a metallic phase. As $U$ increases, a transition to a traveling regime, signifying an Anderson insulator, occurs. Excellent agreement with Eq.~(\ref{TW_eqn}) is found at $U=18.5$, with a velocity of $v\approx -1.21$ (dashed-dotted line). The critical behavior described by \Eq{eq:critsmallsizecorr} emerges at $U_c=12.45$, supported by a numerical fit (dashed line) with parameters $C'=4.3$ and $c_1=2.9$. This consistency between the two methods of determining the transition point highlights the robustness of our results.

\begin{figure}
     \includegraphics[angle=0,width=0.44\linewidth]{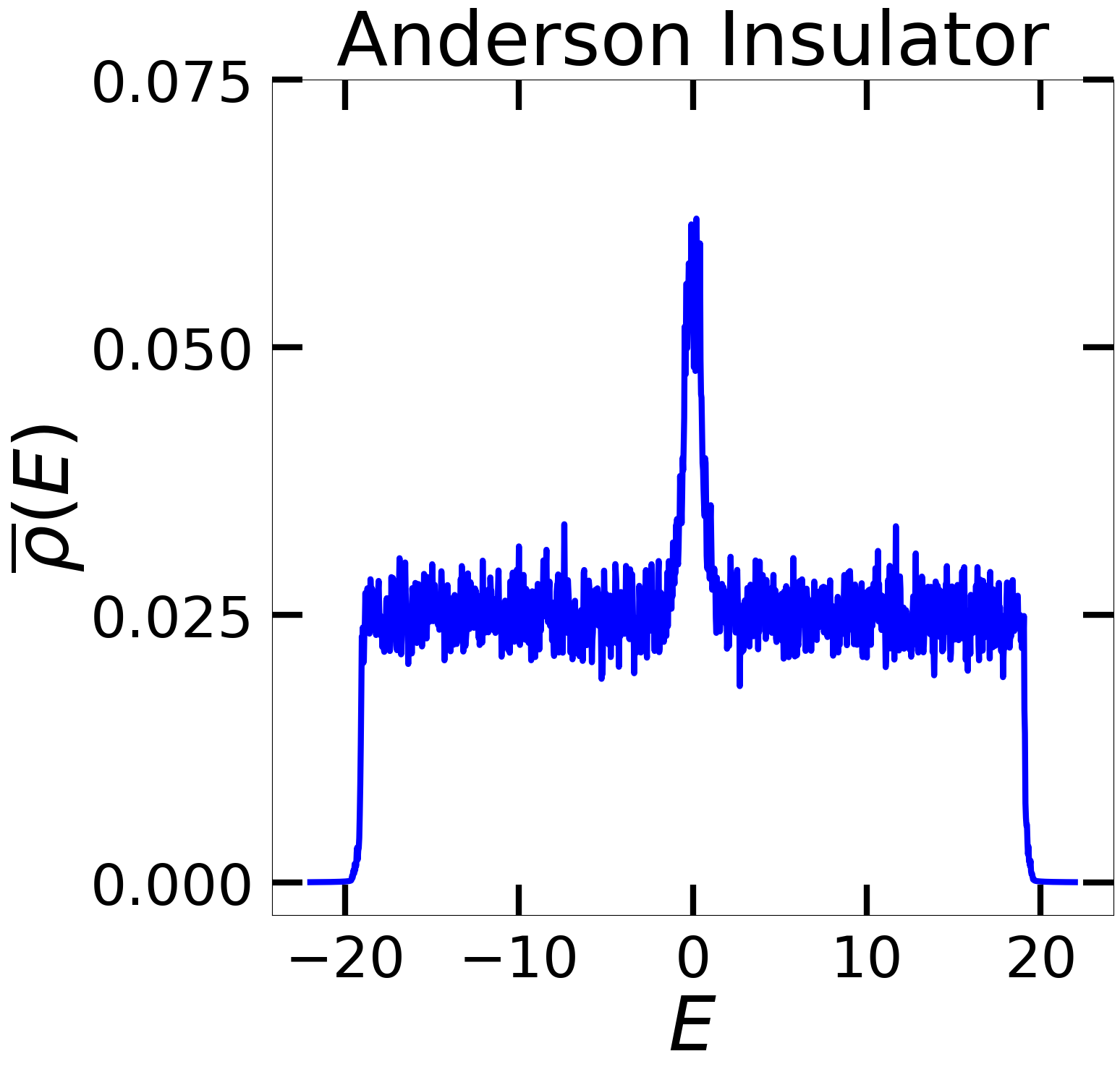}
     \includegraphics[angle=0,width=0.44\linewidth]{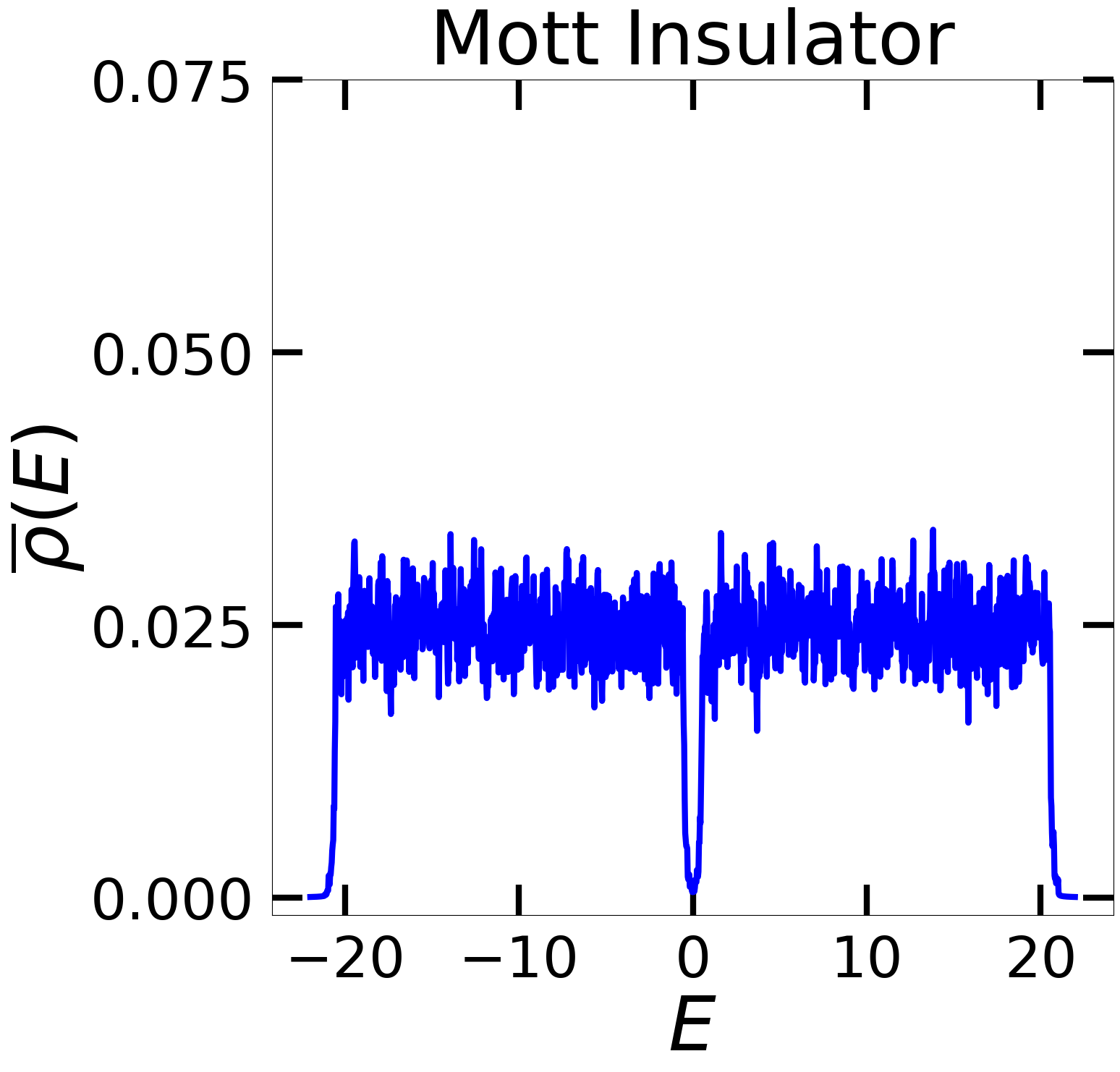} \\
     \includegraphics[angle=0,width=0.95\linewidth]{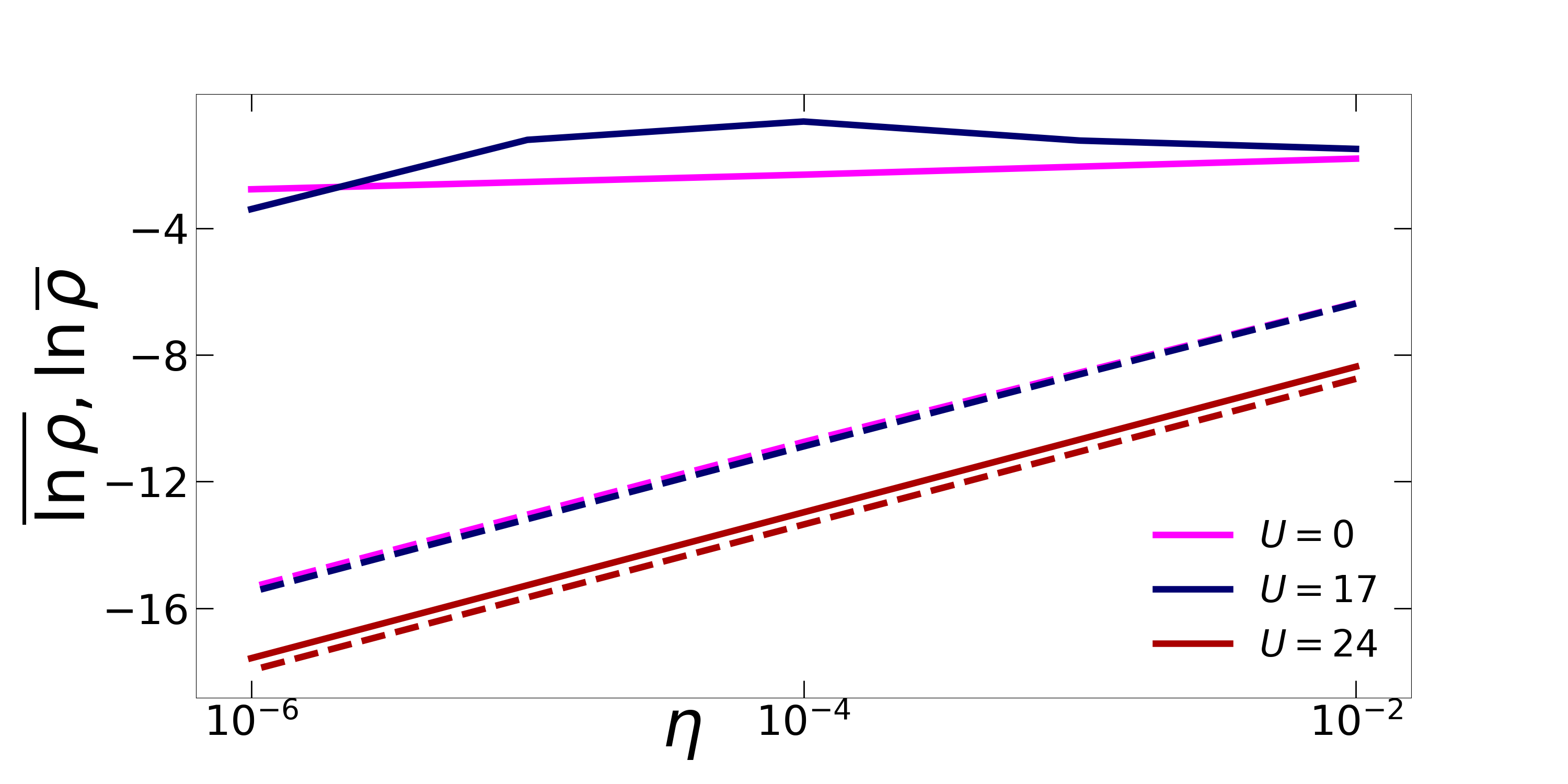}
  \caption{Upper panels: Determination of the Mott transition through the opening of a gap at $E=0$ in the averaged LDOS $\overline{\rho}(E)$. A finite $\eta = 10^{-2}$ at every site, and $n=18$, have been considered. In the left top panel corresponding to $W=20$, $U=19$, we find a non-vanishing $\overline{\rho}(E=0)>0$. A Mott gap opens up for $U\gtrsim W$, as we illustrated in the right top panel for $W=20$ and $U=21$. Lower panel: Distinguishing the Anderson insulator from the Mott insulator based on the behavior of averaged and typical $\rho$ at $E=0$ as a function of $\eta$. In both insulators, the typical value $\rho_{typ}$ decays exponentially with $\eta$ (dashed lines). However, for $U=0$ and $U=17<W$, the averaged value $\overline{\rho}$ (upper magenta and violet lines) is significantly larger than the typical one (dashed lines, same colors) and remains almost independent of $\eta$, characteristic of an Anderson insulating phase. In contrast, for $U=24>W$, $\overline{\rho}$ (lower red line) vanishes exponentially with $\eta$, signifying a Mott gap at $E=0$. }
  \label{Fig_4}
 \end{figure} 

\section{Mott insulator} 

In this final section, we explore the regime of the Mott insulating phase. This involves examining the averaged local density of states, $\overline{\rho}(E)$, as a function of energy $E$. The Mott phase is identified by a gap in $\overline{\rho}(E)$ near $E=0$. Unlike our previous approach with $\eta$ only at the boundary, determining $\overline{\rho}(E)$ accurately necessitates considering a finite $\eta$ \textit{at every site} to mitigate large fluctuations of $\rho$ \cite{kravtsov2018non}. 

Figure~\ref{Fig_4} displays $\overline{\rho}(E)$ for both the Anderson insulator and the Mott insulator. In the former case, like for $W=20$ and $U=19$, non-vanishing $\overline{\rho}(E)$ is observed at the Fermi energy $E=0$. The presence of a Mott gap at the Fermi level is evident, as $\overline{\rho}=0$ for $U=21\gtrsim W=20$. Consequently, the line $U \simeq W$ signifies the boundary between the Anderson insulator and the Mott insulator.

The distinction between the Anderson insulator and the Mott insulator is evident in the behavior of the averaged $\overline{\rho}$ and typical $\rho_{typ}$ at $E=0$ concerning $\eta$, see Fig.~\ref{Fig_4}. In the Anderson insulating phase, the distribution of $\rho$ displays fat tails with $\gamma < 1$, where rare events significantly influence the averaged value, making it much larger than the typical value (which approaches zero): $\overline{\rho} \gg \rho^{typ} (\propto \eta)$ \cite{biroli2012difference, kravtsov2018non}. In contrast, the Mott insulator lacks such substantial fluctuations, leading to $\overline{\rho} \approx \rho^{typ}$, and both tend to zero as $\eta \to 0$.

\section{Conclusion}

In this Letter, we have delved into the metal-insulator transition influenced by disorder and interactions on the Cayley tree, a network characterized by infinite effective dimensionality and deemed ideal for the accuracy of DMFT \cite{DMFT_rev}. However, the pronounced fluctuations of the {local density of states} necessitate the application of statistical DMFT \cite{PhysRevLett.78.3943, dobrosavljevic1998dynamical, SDMFT_3}. Our exploration was prompted by recent findings in the non-interacting regime, revealing non-trivial and strong finite-size effects at the Anderson transition on networks of effective infinite dimensionality \cite{tikhonov2016anderson, pietracaprina2016forward, tarquini2017critical, PhysRevLett.118.166801, tikhonov2019critical, garcia2019two, biroli2018delocalization, Crt_4, sierant2023universality}. These effects complicate the determination of the localization threshold. Notably, recent precise determinations of the threshold (see \cite{tikhonov2019critical, parisi2019anderson, sierant2023universality}, including ours in this study), $W_c^0 \approx 18.17$, significantly surpasses the previously established $W_c^0 \approx 16$ \cite{abou1973selfconsistent}, underscoring the influence of finite-size effects in favoring the metallic regime.

Our investigation revisited the disorder and interaction driven metal-insulator transition on the Cayley tree, challenging previous observations of a direct metal-Mott transition at high disorder \cite{PhysRevLett.78.3943, PhysRevLett.110.066401, PhysRevB.84.115113}. 
An Anderson insulating phase consistently emerges between the metallic and Mott phases. This prediction was meticulously derived by drawing an analogy with a traveling wave problem to describe logarithmic finite-size effects accurately \cite{TW_2, TW_1, monthus2008anderson, chakrabarti2022traveling}. Employing statistical DMFT on a finite Cayley tree with a depth of up to $n=22$, we tackled a very large number of quantum impurity problems. To handle the computational demands, we chose the Hubbard-I approximation \cite{hubbard1963electron} due to its computational efficiency and accuracy in the explored regimes.

A qualitative explanation for the presence of a localized phase between the metal and the Mott insulator can be expressed as follows. At $E=0$, states dwell in the Lifshitz tails of the Mott bands, which are strongly localized at sufficiently large disorder. This phenomenon mirrors the Superfluid to Mott insulator transition in disordered Bosons, where a Bose glass, a localized insulating phase, consistently separates the Superfluid and Mott phases \cite{PhysRevLett.103.140402, PhysRevB.80.214519} (see \cite{dupont2020dirty} for the Cayley tree case).

An intriguing open question arises concerning the nature of the metallic phase. In the non-interacting limit, the delocalized phase exhibits non-ergodicity on the Cayley tree, characterized by multifractal properties \cite{monthus2011anderson, deluca2014anderson, tikhonov2016fractality, kravtsov2018non, biroli2018delocalization}. With interactions, this scenario might give rise to a non-Fermi liquid (Griffiths) metallic phase, previously discussed in the literature \cite{PhysRevLett.78.3943, andrade2009electronic, PhysRevLett.110.066401}. It is noteworthy that this phase was previously found just before the Mott phase; however, our prediction suggests an Anderson insulator instead. On the Cayley tree, the whole metallic phase would manifest as a Griffiths phase in this scenario.

Exploring the regime of weak disorder presents an intriguing perspective. A distinctive trait in this regime, in the non-interacting case, is the delocalization of Lifshitz states at sufficiently weak disorder, see \cite{PhysRevLett.106.136804}. This scenario suggests a direct transition from metal to Mott, aligning with previous observations \cite{PhysRevLett.78.3943, PhysRevLett.110.066401, PhysRevB.84.115113}. To delve into this limit, 
surpassing the constraints of the Hubbard-I approximation used in our study becomes imperative. 
Additionally, the finite-dimensional case offers another avenue for exploration, potentially addressable through the forward scattering approximation \cite{pietracaprina2016forward}, which could ultimately pave the way to a first-principle description of correlated materials where disorder is expected to play a major role, such as layered dichalcogenide 1T-TaS$_2$ with Cu-intercalation \cite{PhysRevLett.112.206402} or doped Ru- or Ir-based oxides\cite{PhysRevB.71.125104,PhysRevB.92.081114,PhysRevB.86.125105}.

\begin{acknowledgments}
	We thank J. Gong and A. Berger for interesting discussions.
	This study has been supported by the French National Research Agency  (ANR) under  projects COCOA ANR-17-CE30-0024, MANYLOK  ANR-18-CE30-0017 and GLADYS ANR-19-CE30-0013, the   EUR  grant NanoX  No. ANR-17-EURE-0009  in  the  framework  of the ``Programme des Investissements d'Avenir'', and by the Singapore Ministry of Education Academic Research Fund Tier I (WBS No. R-144-000-437-114). 
 This work was granted access to the HPC resources of CALMIP and the National Supercomputing Centre (NSCC), Singapore.
\end{acknowledgments}

\bibliographystyle{eplbib.bst}
\bibliography{AHM_Cayley}

\end{document}